\documentclass[a4paper]{spie}  

\usepackage{amsmath,amsfonts,amssymb}
\usepackage{graphicx}
\usepackage[colorlinks=true, allcolors=blue]{hyperref}

\title{Progress on the UV-VIS arm of SOXS}

\author[a*]{Adam Rubin}
\author[b]{Sagi Ben-Ami}
\author[b]{Ofir Hershko}
\author[b]{Michael Rappaport}
\author[b]{Avishay Gal-Yam}
\author[b]{Rachel Bruch}
\author[c]{Sergio Campana}
\author[d]{Riccardo Claudi}
\author[e]{Pietro Schipani}
\author[c]{Matteo Aliverti}
\author[d]{Andrea Baruffolo}
\author[f]{Federico Biondi}
\author[e]{Giulio Capasso}
\author[g,h]{Rosario Cosentino}
\author[i]{Francesco D'Alessio}
\author[c]{Paolo D'Avanzo}
\author[j,k]{Hanindyo Kuncarayakti}
\author[c]{Marco Landoni}
\author[h]{Matteo Munari}
\author[l,m]{Giuliano Pignata}
\author[n,h]{Salvatore Scuderi}
\author[i]{Fabrizio Vitali}
\author[o]{David Young}
\author[p]{Jani Achrén}
\author[q,m]{José Antonio Araiza-Duran}
\author[r]{Iair Arcavi}
\author[l,s]{Anna Brucalassi}
\author[d]{Enrico Cappellaro}
\author[e]{Mirko Colapietro}
\author[e]{Massimo Della Valle}
\author[d]{Marco De Pascale}
\author[h]{Rosario Di Benedetto}
\author[e]{Sergio D'Orsi}
\author[t]{Thomas Flügel-Paul}
\author[c]{Matteo Genoni}
\author[g]{Marcos Hernandez}
\author[k,j]{Jari Kotilainen}
\author[u]{Gianluca Li Causi}
\author[j]{Seppo Mattila}
\author[d]{Kalyan Radhakrishnan}
\author[d]{Davide Ricci}
\author[c]{Marco Riva}
\author[t]{Susann Sadlowski}
\author[d]{Bernardo Salasnich}
\author[o]{Stephen Smartt}
\author[h]{Ricardo Zanmar Sanchez}
\author[v]{Maximilian Stritzinger}
\author[g]{Hector Ventura}

\affil[a]{European Southern Observatory, Garching bei München, Germany}
\affil[b]{Weizmann Institute of Science, Rehovot, Israel}
\affil[c]{INAF - Osservatorio Astronomico di Brera, Merate, Italy}
\affil[d]{INAF - Osservatorio Astronomico di Padova, Padua, Italy}
\affil[e]{INAF - Osservatorio Astronomico di Capodimonte, Naples, Italy}
\affil[f]{Max-Planck-Institut für Extraterrestrische Physik, Garching, Germany}
\affil[g]{INAF - Fundación Galileo Galilei, Breña Baja, Spain}
\affil[h]{INAF - Osservatorio Astrofisico di Catania, Catania, Italy}
\affil[i]{INAF - Osservatorio Astronomico di Roma, Rome, Italy}
\affil[j]{Tuorla Observatory, Department of Physics and Astronomy, University of Turku, Turku, Finland}
\affil[k]{FINCA - Finnish Centre for Astronomy with ESO, Turku, Finland}
\affil[l]{Universidad Andres Bello, Santiago, Chile}
\affil[m]{Millennium Institute of Astrophysics (MAS), Santiago, Chile}
\affil[n]{INAF - Istituto di Astrofisica Spaziale e Fisica Cosmica, Milano, Italy}
\affil[o]{Queen's University Belfast, Belfast, UK}
\affil[p]{Incident Angle Oy, Turku, Finland}
\affil[q]{Centro de Investigaciones en Optica A. C., León, Mexico}
\affil[r]{Tel Aviv University, Tel Aviv, Israel}
\affil[s]{INAF-Osservatorio Astrofisico Arcetri, Firenze, Italy}
\affil[t]{Fraunhofer-Institut für Angewandte Optik und Feinmechanik IOF, Jena, Germany}
\affil[u]{INAF - Istituto di Astrofisica e Planetologia Spaziali, Rome, Italy}
\affil[v]{Aarhus University, Aarhus, Denmark}

\authorinfo{*arubin@eso.org}

\begin{document} 
\maketitle

\begin{abstract}
We present our progress on the UV-VIS arm of Son Of X-Shooter (SOXS), a new spectrograph for the NTT. Our design splits the spectral band into four sub-bands that are imaged onto a single detector. Each band uses an optimized high efficiency grating that operates in 1st order (m=1). In our previous paper we presented the concept and preliminary design. SOXS passed a Final Design Review in July 2018 and is well into the construction phase. Here we present the final design, performances of key manufactured elements, and the progress in the assembly. Based on the as-built elements, the expected throughput of the visual arm will be $>55\%$. This paper is accompanied by a series of contributions describing the progress made on the SOXS instrument.
\end{abstract}

\keywords{spectrograph, transients}

\section{INTRODUCTION}
\label{sec:intro}  
The Son Of X-Shooter (SOXS) \cite{schipani_new_2016,schipani_soxs_2018,schipani_development_2020} is a medium resolution spectrograph ($R\sim4500$ for a 1 arcsecond slit) proposed for the ESO 3.6 m NTT. SOXS is comprised of a common path, which splits the incoming light to a UV-VIS arm ($0.35-0.85$ $\mu$m) and a NIR arm ($0.8-2.0$ $\mu$m), an acquisition camera, and a calibration unit---all of which are described in other papers in these proceedings.\cite{aliverti_manufacturing_2020,biondi_aiv_2020,brucalassi_final_2020,claudi_operational_2020,colapietro_progress_2020,cosentino_development_2020,genoni_soxs_2020,kuncarayakti_design_2020,ricci_development_2020,sanchez_soxs_2020,schipani_development_2020,vitali_development_2020,young_soxs_2020}

\begin{figure}
    \centering
    \includegraphics[width=0.5\textwidth]{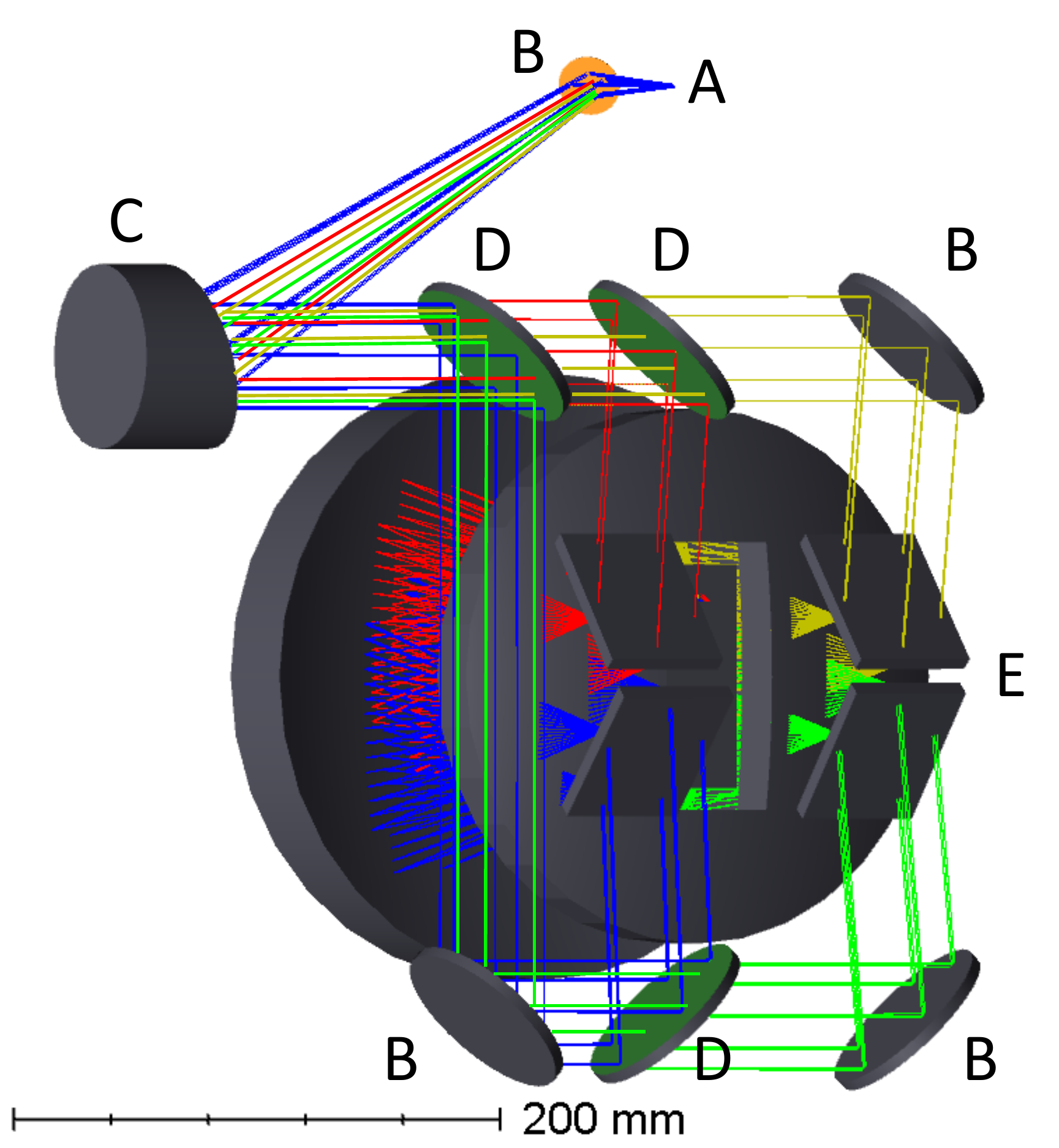}
    \caption{System layout: A. slit plane, B. Reflective mirrors, C. OAP Collimator, D. Dichroic filters, E. Gratings. }
    \label{fig:sys_layout}
\end{figure}

\begin{figure}
    \centering
    \includegraphics[width=0.45\textwidth]{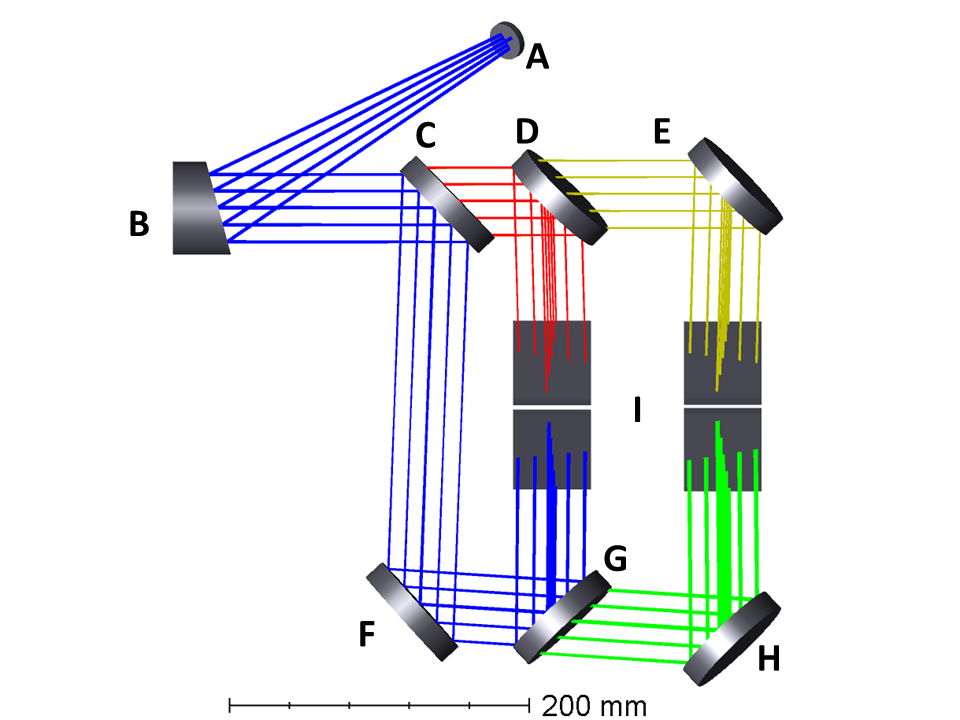}
    \includegraphics[width=0.45\textwidth]{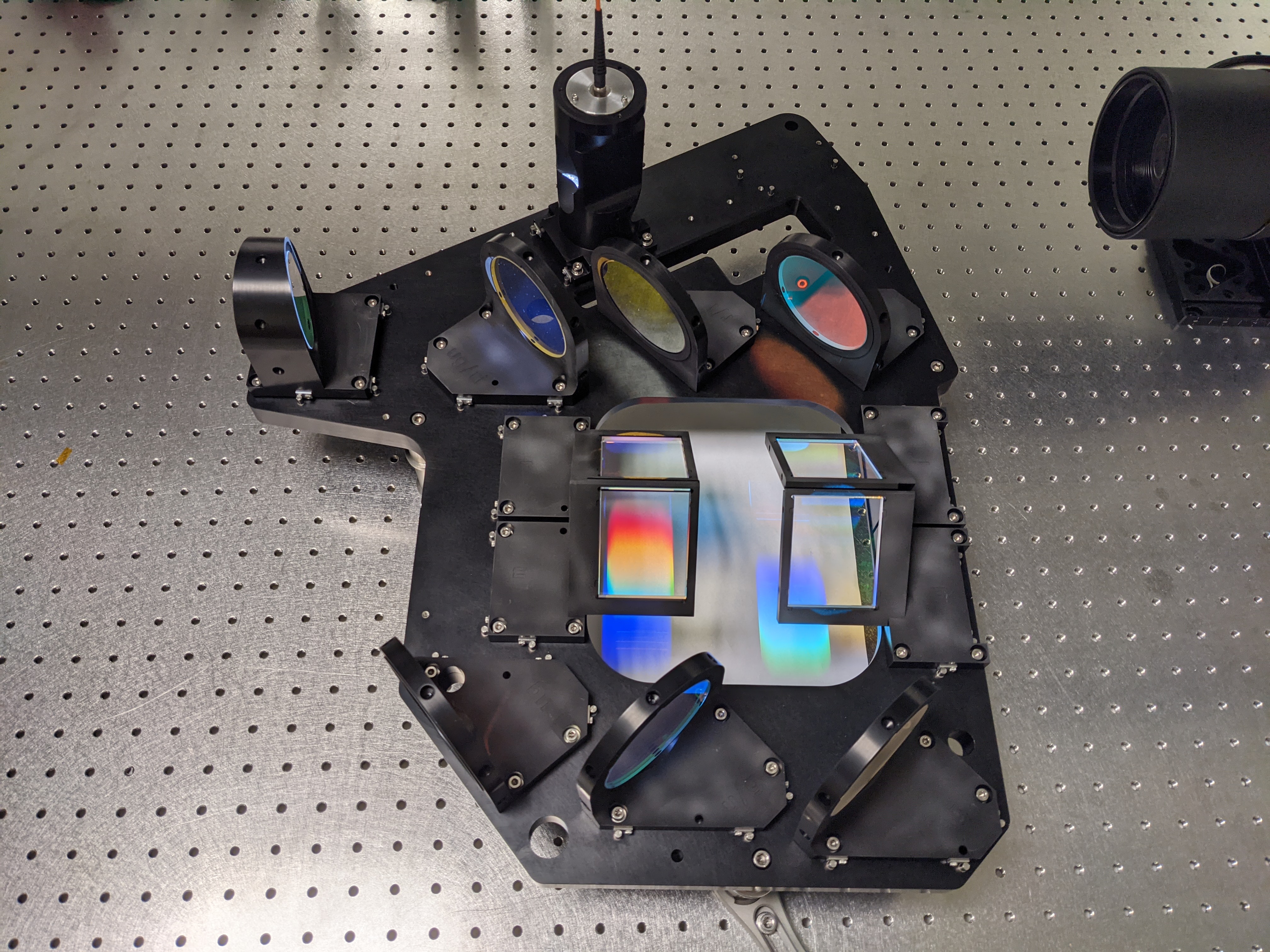}
    \caption{Left: Layout of the feed. A: Folding mirror; B: Off-axis parabola collimator; C/D/G: Dichroic mirrors dividing $u+g$ from $r+i$, $r$ from $i$, and $u$ from $g$, respectively; E/F/H: Dielectric mirrors reflecting $i$, $u+g$, and $g$, respectively; I: gratings. Right: the as-built feed illuminated withe a white light source.}
    \label{fig:feed_labels}
\end{figure}

\begin{figure}
    \centering
    \includegraphics[width=0.65\textwidth]{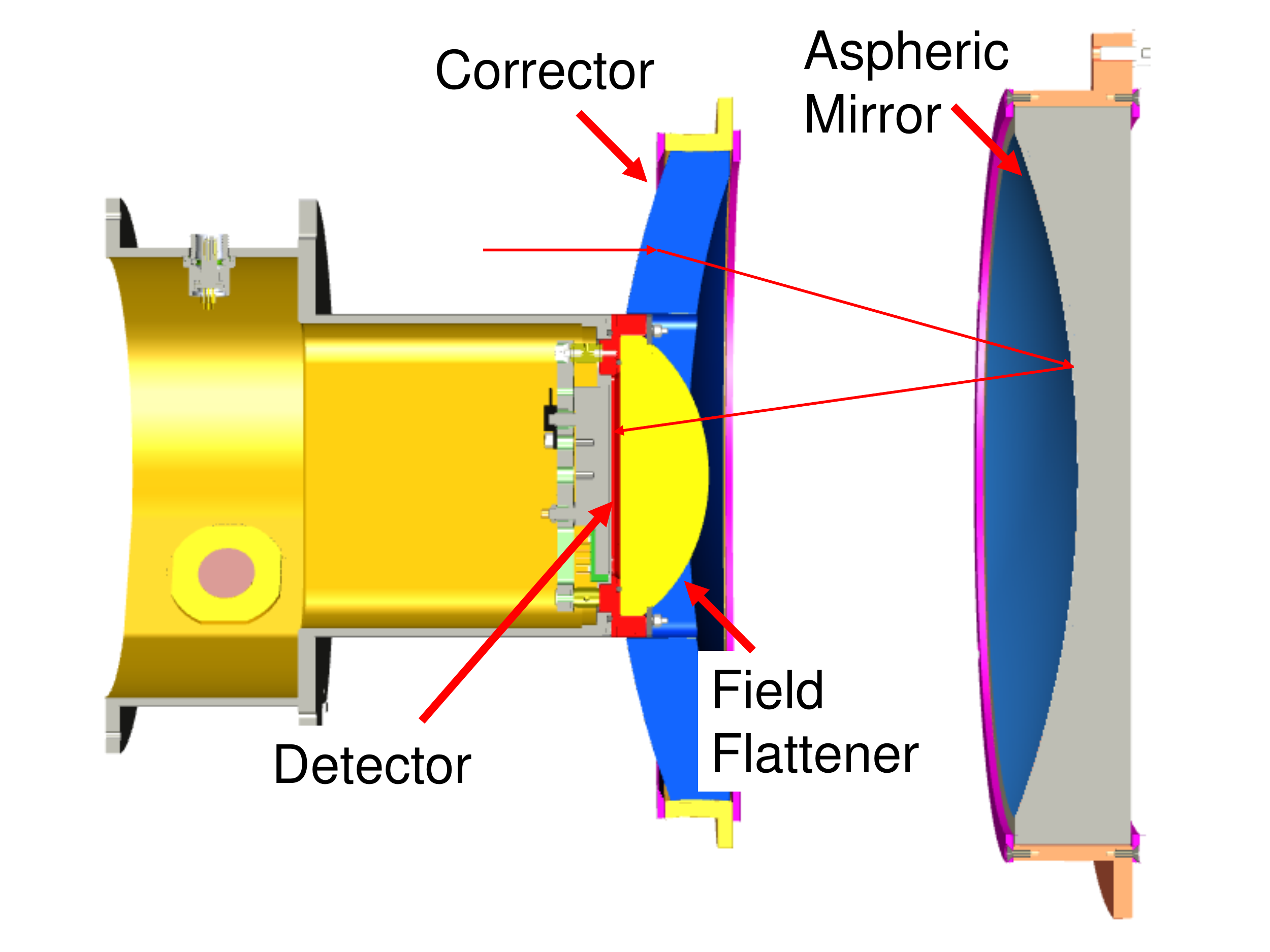}
    \caption{Camera layout. An illustration of the light path is shown in red.}
    \label{fig:cam_layout}
\end{figure}

The visual arm has a novel design, which was described in our previous paper.\cite{rubin_mits:_2018} The beam is divided into four bands and imaged by a single camera. Each band makes use of a high efficiency grating in first order $(m=1)$ to increase the photon efficiency. The optical layout is shown in Figure \ref{fig:sys_layout}. The UV-VIS spectrograph has two separated levels, one called the ``feed'', where the source image is collimated, split with dichroics and mirrors, and dispersed by the gratings. The feed layout is shown in Figure \ref{fig:feed_labels}. The dispersed beam is directed to the lower stage which is comprised of a single camera which images all beams onto one detector. The camera is made up of a CaF2 corrector, a fused silica mirror and a fused silica field flattener---each having one aspheric surface. The camera layout is shown in Figure \ref{fig:cam_layout}.

In the period between the previous paper and this manuscript the optics---gratings, mirrors, filters, collimator and camera---have been manufactured. Most of the mechanical structure has been manufactured, and the continuous flow cryostat and vacuum vessel are currently being tested. The system is now in the assembly, integration and verification phase with work ongoing. The entire SOXS system is currently scheduled to undergo preliminary acceptance in Europe by the end of 2021.

\section{As-built performance of elements}
\subsection{Gratings}
\begin{figure}
    \centering
    \includegraphics[width=0.75\textwidth]{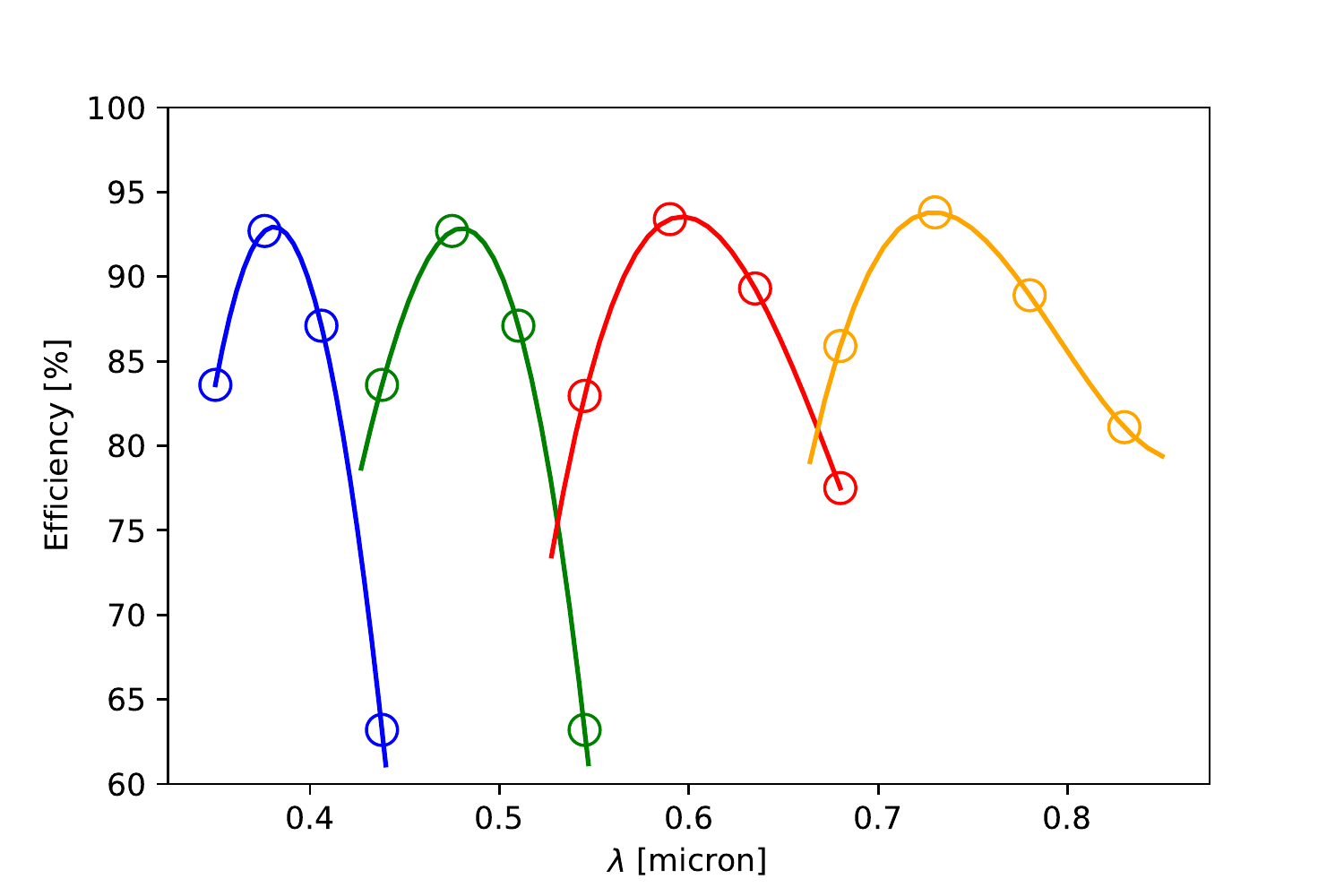}
    \caption{SOXS visual arm as-built grating efficiencies as measured by the manufacturer. The average between the polarizations is shown. The lines are a third-order polynomial fit to the data.}
    \label{fig:grating_eff}
\end{figure}

\begin{figure}
    \centering
    \includegraphics[width=0.45\textwidth]{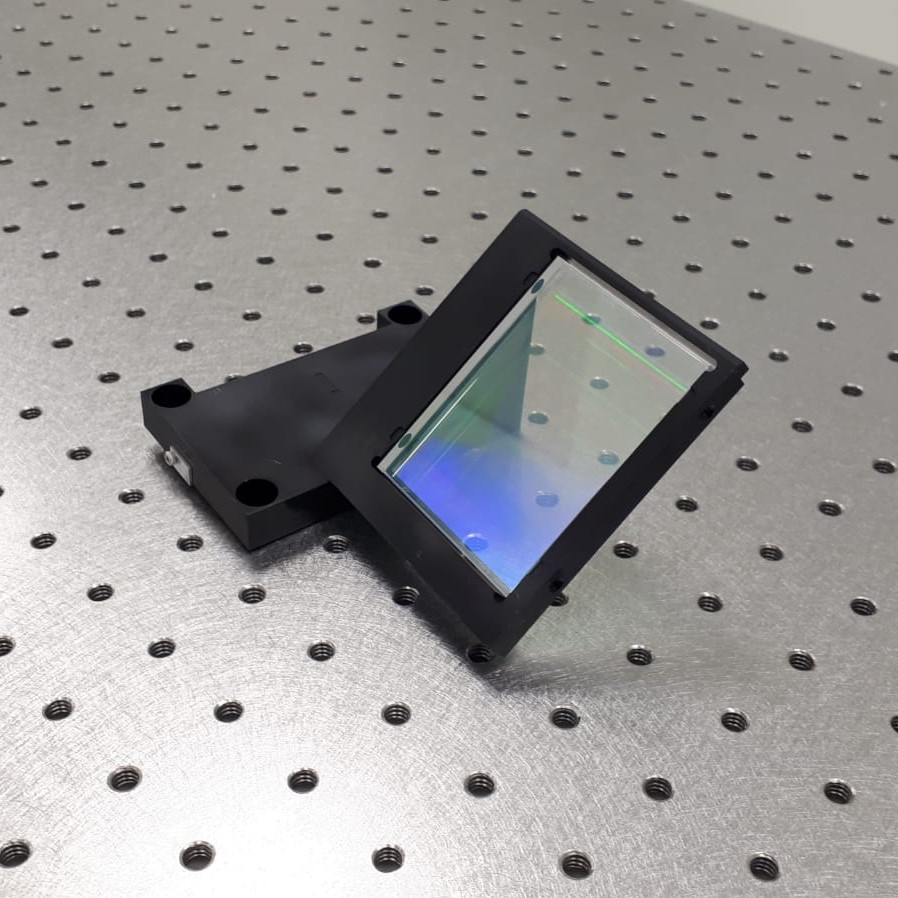}
    \includegraphics[width=0.45\textwidth]{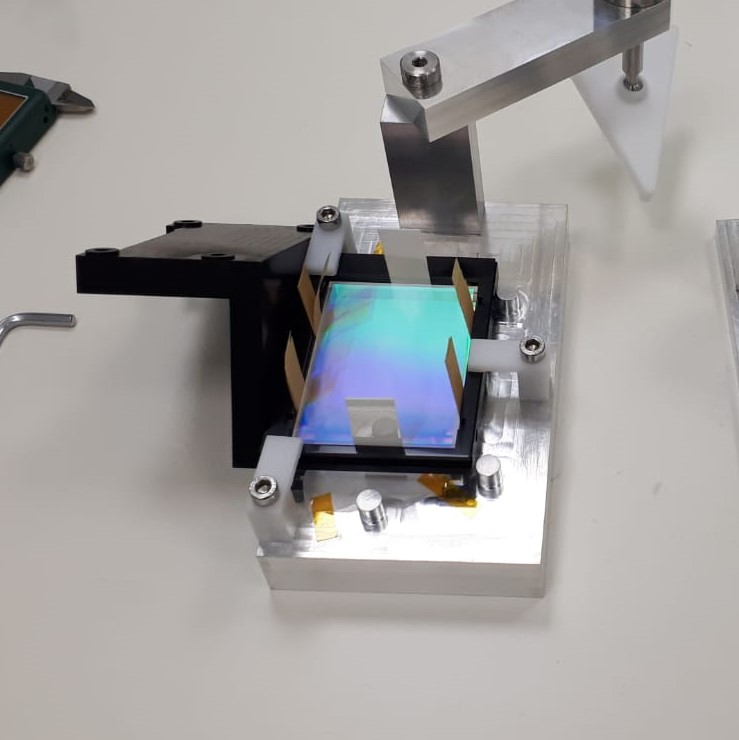}
    \caption{Grating adhered in its mount (left) and in the gluing jig (right). Shims were used to center the element during the gluing process and to ensure the correct thickness of RTV at each adhesion point.}
    \label{fig:grating_pics}
\end{figure}

The four gratings for the SOXS visual arm, produced by using atomic layer deposition on fused silica, were manufactured by Fraunhofer IOF, Jena. Their parameters are given in Table \ref{tab:grating_params}. Figure \ref{fig:grating_eff} shows the average efficiency (between TM and TE polarizations) for the four gratings. Figure \ref{fig:grating_pics} shows the u-band grating in its mount and gluing jig. The general procedure for mounting the small optics into their mounts is describe in section \ref{sec:mounting}.

\begin{table}
	\centering
	\caption{SOXS UV-VIS and grating parameters.\label{tab:grating_params}}
	\begin{tabular}{|c|c|c|c|c|}
	\hline
	Quasi-Order & Wavelength Range & Line Density & AOI & $\lambda_{Littrow}$ \\
	& [nm] & [lines/mm] &  &[nm] \\ \hline 
	$u$ & $350-440$ & 3378.4 & $41^\circ$ & 388.4 \\ \hline
	$g$ & $427-547$ & 2652.5 & $41^\circ$ & 494.7 \\ \hline
	$r$ & $527.5-680$ & 2118.6 & $41^\circ$ & 619.3 \\ \hline
	$i$ & $664-850$ & 1709.4 & $41^\circ$ & 767.6	\\ \hline 
	\end{tabular}
\end{table}

\subsection{Off-axis parabola}

The off-axis parabola collimator was manufactured by Winlight System. The parabola has a 30 degree off-axis angle and a design radius of $545.11$ mm. The parabola was manufactured with within $\pm0.015\%$ of the design radius and with an RMS surface figure error of $\sim20$ nm.

\begin{figure}
    \centering
    \includegraphics[width=0.55\textwidth]{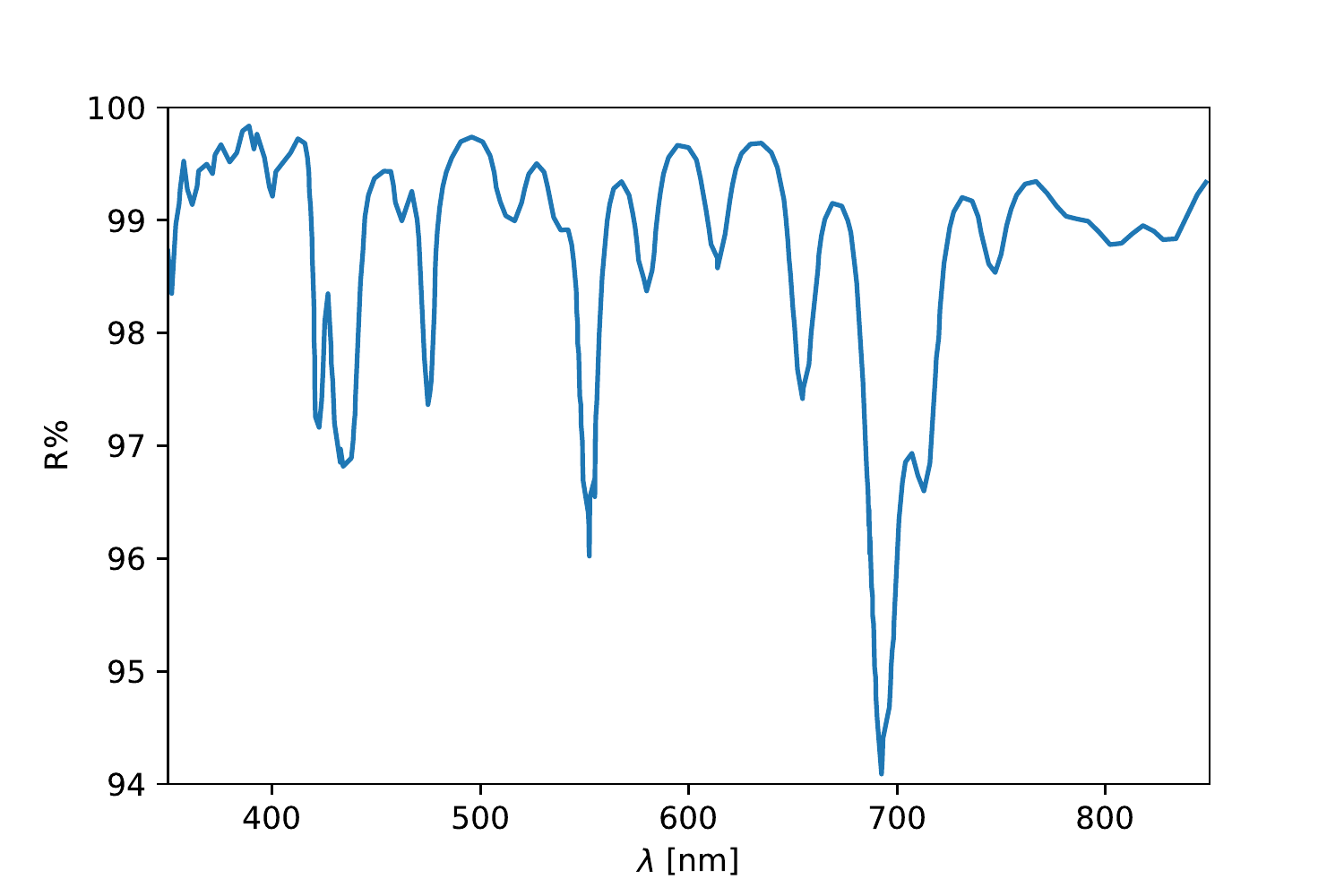}
    \caption{Measured reflection efficiency of the off-axis parabola.}
    \label{fig:oap_eff}
\end{figure}

\begin{figure}
    \centering
    \includegraphics[width=0.45\textwidth]{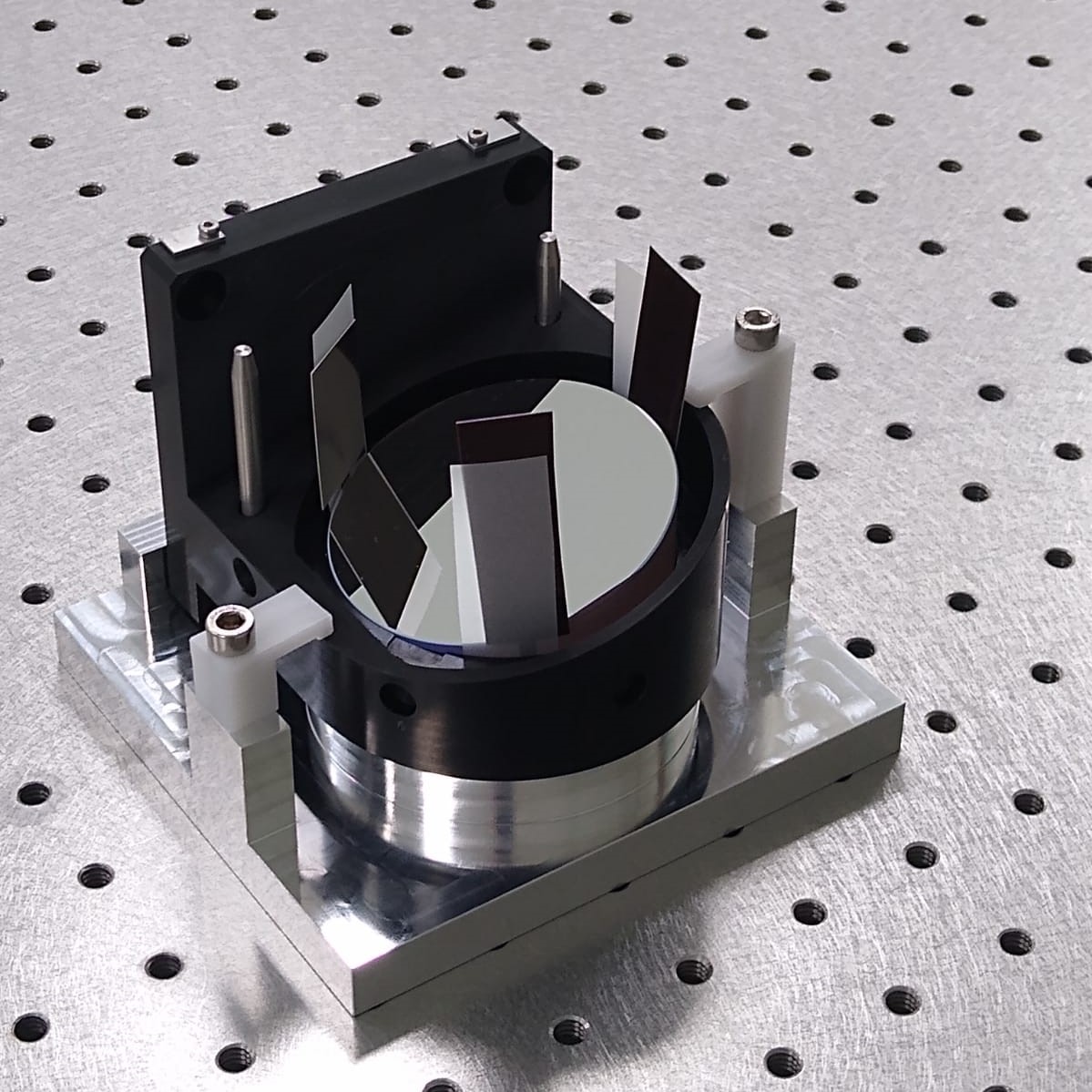}
    \caption{Off-axis parabola in its gluing jig.}
    \label{fig:oap_pics}
\end{figure}

\subsection{Mirrors and dichroic filters}

\begin{figure}
    \centering
    \includegraphics[width=0.75\textwidth]{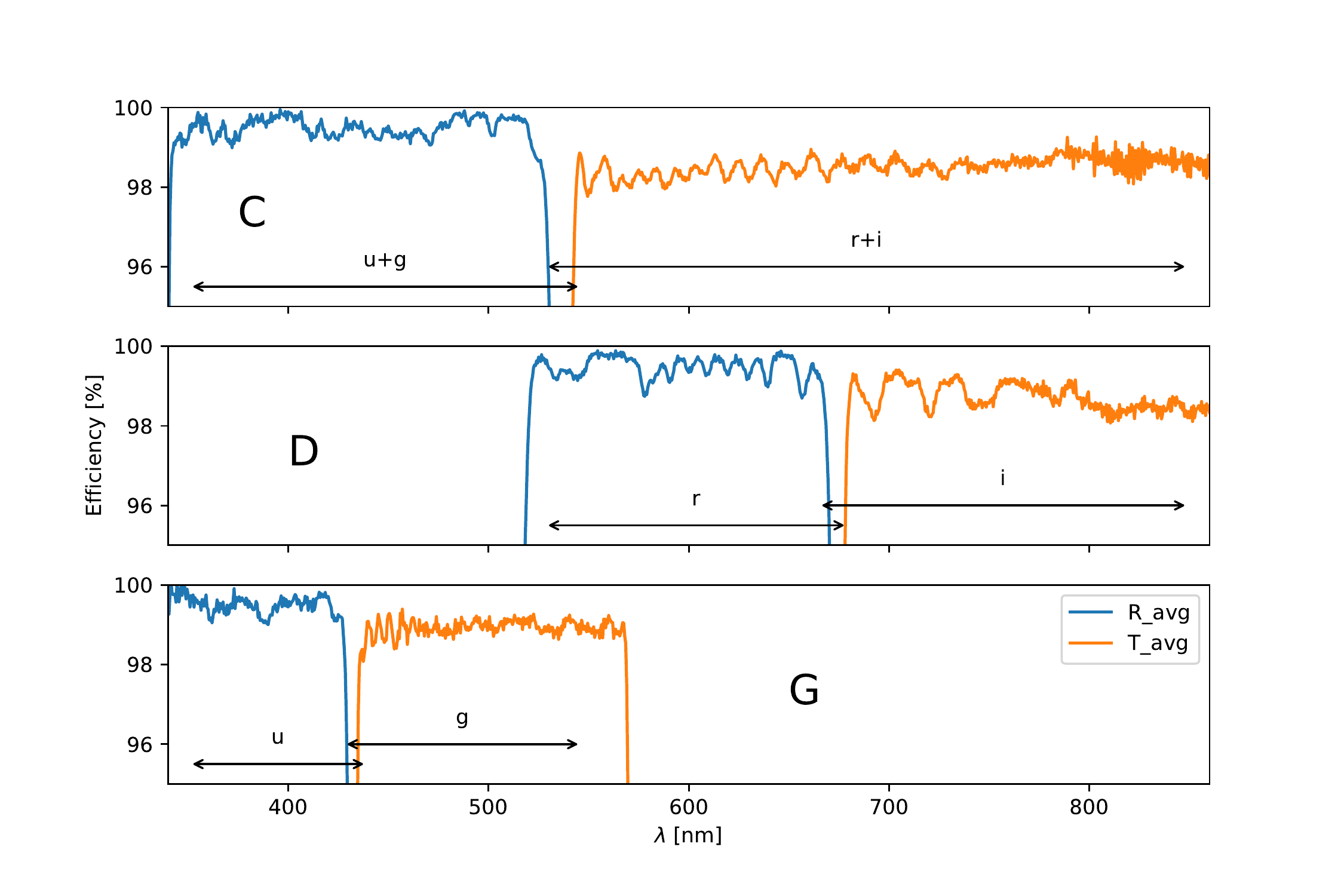}
    \caption{From the top: transmission and reflection of the dichroic filter which splits the light between the $u+g$ and $r+i$ bands, between the $u$ and $g$ bands, and between the $r$ and $i$ bands, as measured by the manufacturer. Arrows mark the region that falls on the detector in each band. The plots are labeled according to the labels in Figure \ref{fig:feed_labels}.}
    \label{fig:dich-ug-ri}
\end{figure}

\begin{figure}
    \centering
    \includegraphics[width=0.75\textwidth]{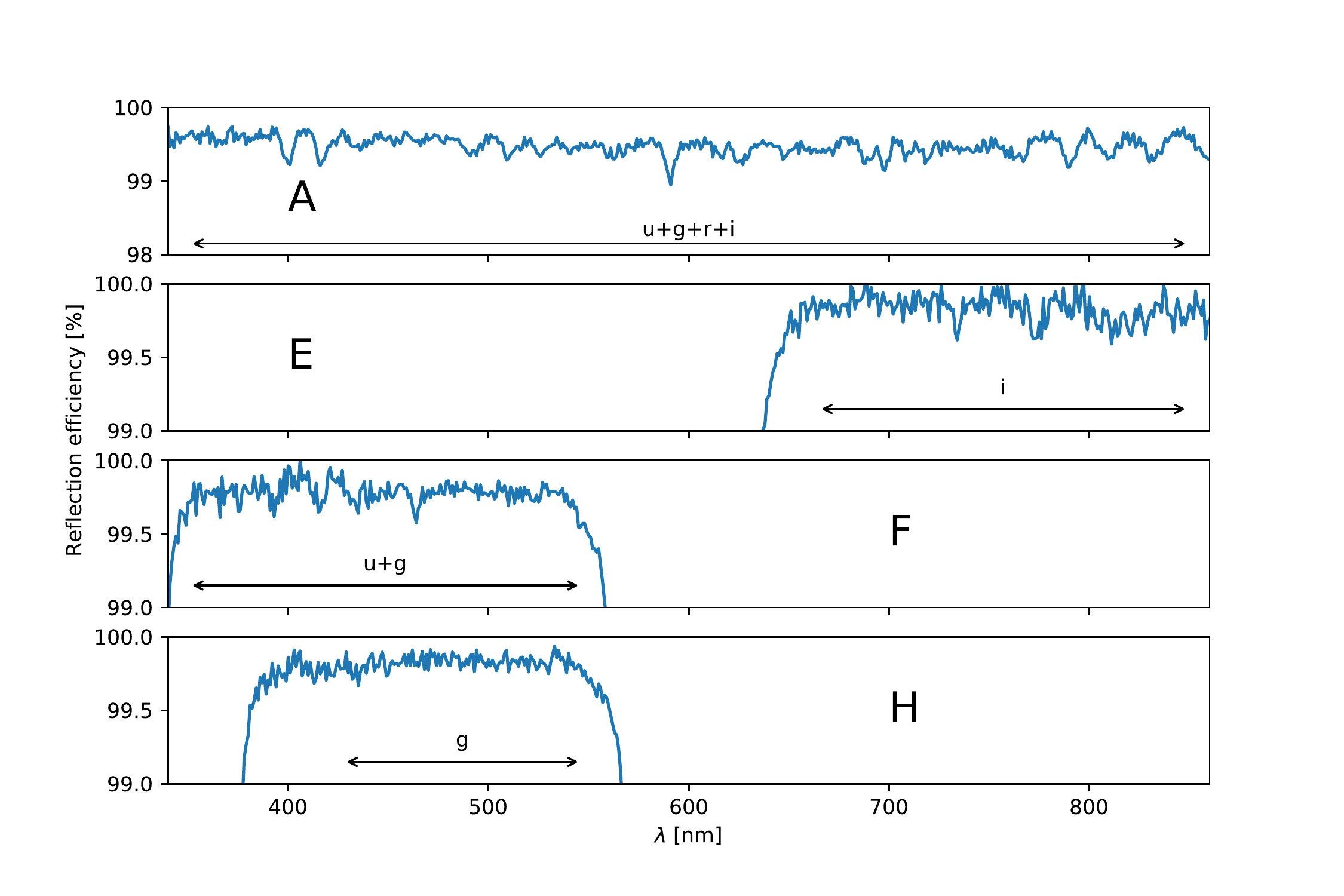}
    \caption{Measured reflection efficiencies for the feed mirrors manufactured by Asahi. Arrows mark the region that falls on the detector in each band. The plots are labeled according to the labels in Figure \ref{fig:feed_labels}.}
    \label{fig:mirrors_eff}
\end{figure}

The flat mirrors and dichroic filters used in the feed were manufactured by Asahi Spectra to $\lambda/8$ flatness. The mirrors and filters were measured by the manufacturer at their design angle of incidence. The average reflectance for all was $>99\%$ and transmission for the filters was measured to be $>98\%$. One of the most challenging elements was element A (fold mirror toward the collimator, see Figure \ref{fig:feed_labels}). This element reflects the entire spectrum, and receives an incident F/6.5 beam at $45^\circ$. As a result, the requirements were for good spectral reflectivity over the entire spectrum for angles of incidence ranging from 41 to 49 degrees. The element was measured under these conditions and satisfied average reflectance above 99\% over the relevant spectral range and angles of incidence. For the full response of each of the elements, see Figures \ref{fig:dich-ug-ri} and \ref{fig:mirrors_eff}.

\begin{table}[ht]
    \centering
    \begin{tabular}{|l|l|l|l|l|l|}
\hline     Element	&	Diameter [mm]	&	Reflected [nm]	&	Transmitted [nm]	&	Transition [nm]	\\
\hline A	&	25.4   &	$350-850	$&	-	&	-	\\
\hline C	&	76.2	&$	350-527.5$&		$547-850$	&	$527.5-547$	\\
\hline D	&	76.2	&$	527.5-664$&		$680-850$	&	$664-680$	\\
\hline E	&	76.2	&$	664-850	$&	-	&	-	\\
\hline F	&	76.2	&$	350-547	$&	-	&	-	\\
\hline G	&	76.2	&$	350-427	$&	$440-547$	&	$427-440$	\\
\hline H	&	76.2	&$	427-547	$&	-	&	-	\\
\hline
    \end{tabular}
    \caption{Specifications for the filters and mirrors manufactured by Asahi. Element labels match the ones that appear in Figure \ref{fig:feed_labels}.}
    \label{tab:asahi_mirrors}
\end{table}

\subsection{Camera}

\begin{figure}
    \centering
    \includegraphics[width=0.75\textwidth]{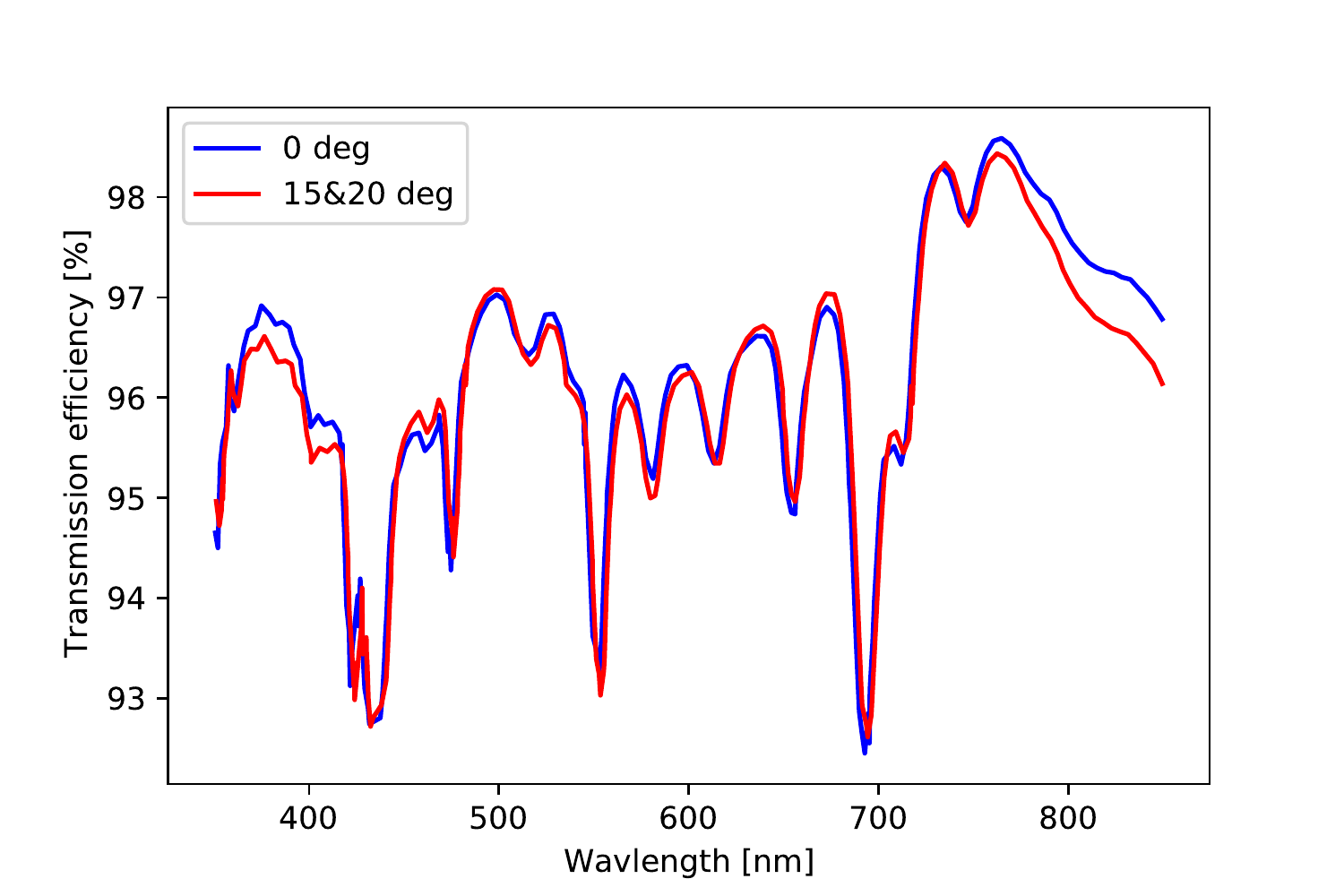}
    \caption{Measured transmission for the entire camera: corrector, mirror, and field flattener. The average efficiency is $>96\%$.}
    \label{fig:cam_eff}
\end{figure}

The camera optics are manufactured by Winlight System. The camera is comprised of a CaF2 corrector, a fused silica mirror, and a fused silica field flattener---each with one aspheric surface. The field flattener serves as the detector window. The camera layout is shown in Figure \ref{fig:cam_layout}. The elements were manufactured to tolerance, with excellent RMS figure error of $<30$ nm, and roughness $<1$ nm. As of the writing of this manuscript, the corrector is being re-coated due to a manufacturing process error. However, witness sample measurements have already been performed and a good estimate of the total throughput of the camera can be estimated. The camera has an average photon efficiency of $>96\%$ and is shown in Figure \ref{fig:cam_eff}.

\section{Mechanical structure}

The mechanical structure and mounts for the feed and small optics (filters, mirrors and gratings) have been manufactured. The Invar truss which holds the camera optics is in advanced stages of production.

\begin{figure}
    \centering
    \includegraphics[width=0.75\textwidth]{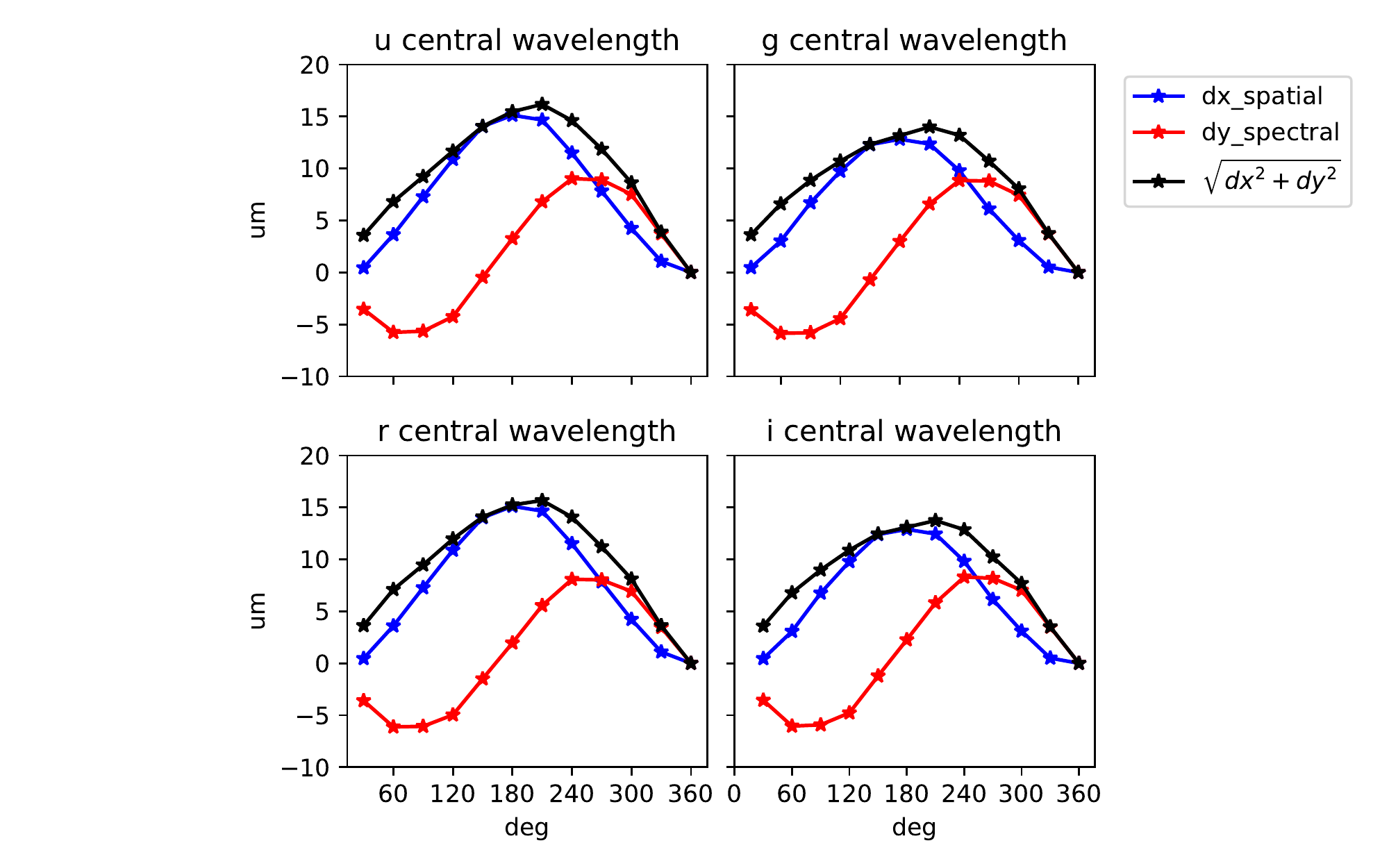}
    \caption{Shift of the spectrum on the detector due to rotation of the instrument on the Nasmyth flange.}
    \label{fig:flexure}
\end{figure}

With the final design in hand, a flexure analysis showed that we expect less than one pixel motion due to flexure of the UV-VIS spectrograph. SOXS will be mounted on the Nasmyth flange of the NTT. Therefore it is expected to rotate during observations, inducing a strongly varying gravity load. We performed a flexure analysis to estimate the displacement of the spectrum at different instrument orientations. We performed a finite element analysis in COMSOL Multiphysics\textsuperscript{\textregistered}. Each optical element's tilt and displacement were recorded and implemented in Optics Studio. Then the position of the central wavelength was recorded for each orientation of the instrument. The results are shown in Figure \ref{fig:flexure}. Figure \ref{fig:flexure} shows that for an entire 360 degree rotation the spectrum moves by a total of one pixel. However this scenario requires a long exposure while pointing very close to the zenith and is should occur very rarely.

\begin{figure}
    \centering
    \includegraphics[width=0.45\textwidth]{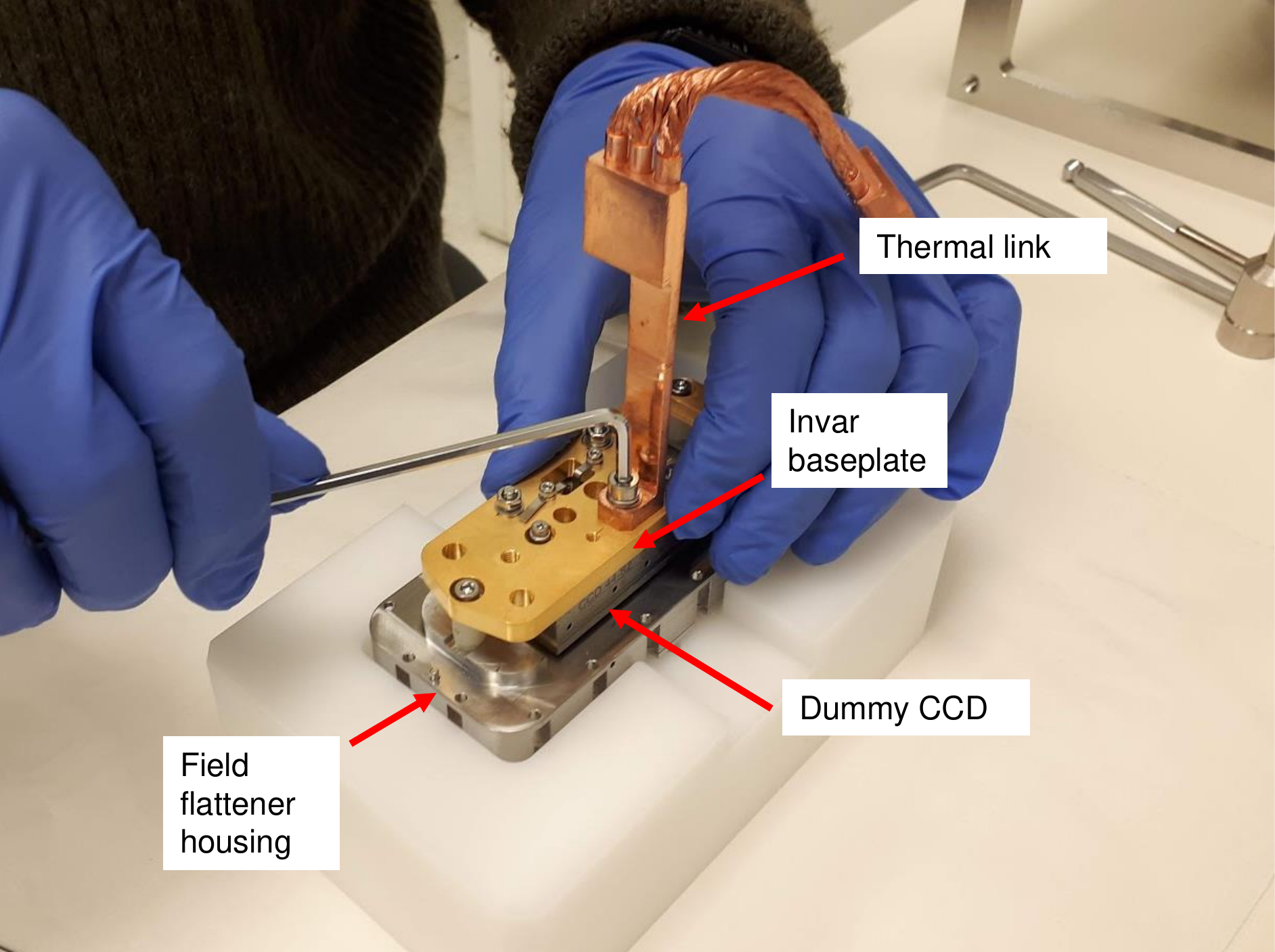}
    \caption{Field flatenner housing, dummy CCD, Invar baseplate, and copper thermal link.}
    \label{fig:ccd_baseplate}
\end{figure}

\begin{figure}
    \centering
    \includegraphics[width=0.45\textwidth]{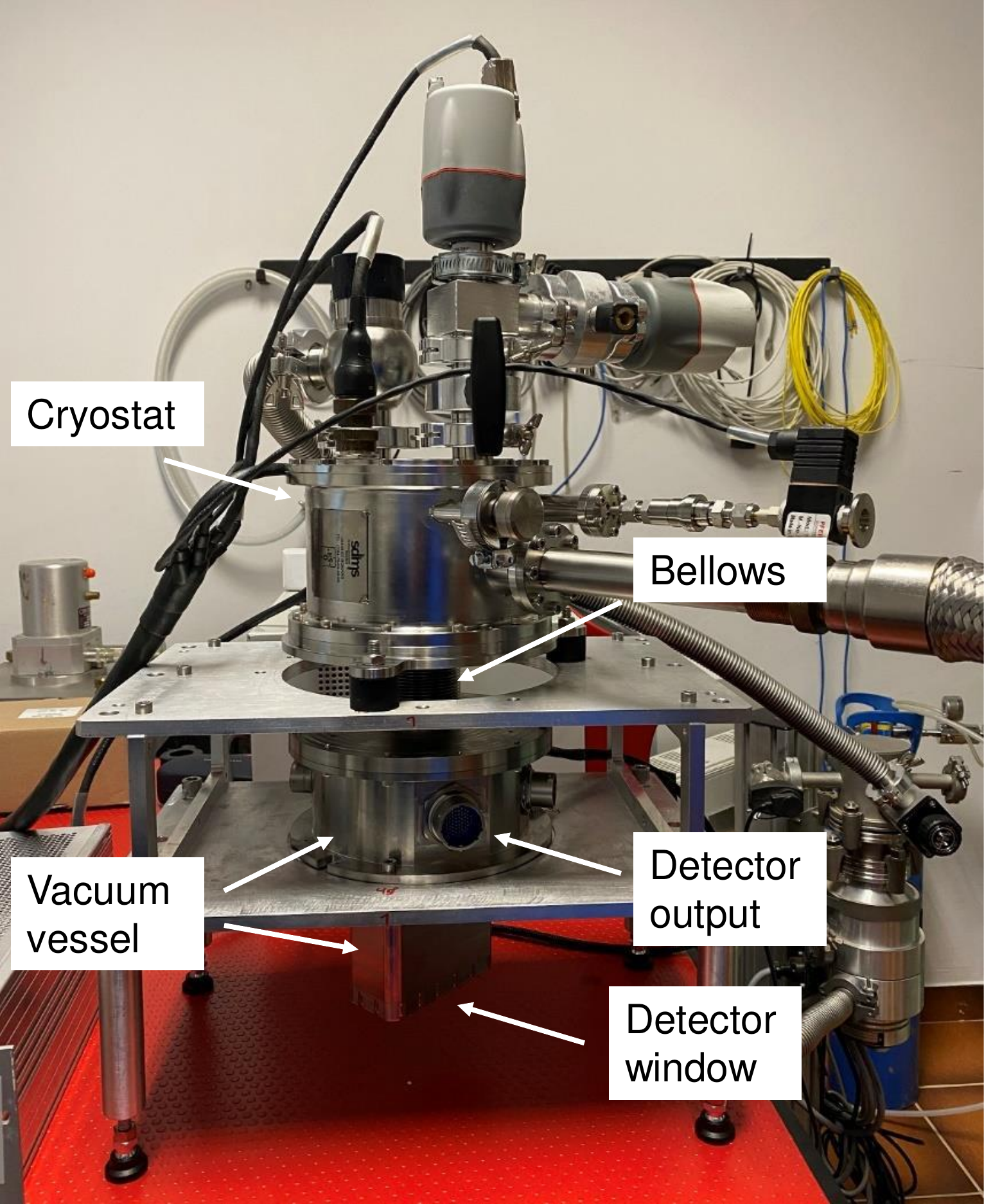}
    \caption{Vacuum chamber and cryostat. A welded bellows separates the cryostat from the vacuum vessel mechanically, preventing flexure due to the weight of the cryostat.}
    \label{fig:vacuum}
\end{figure}

The vacuum chamber and detector mount were manufactured at the Weizmann Institute. The chamber is internally gold plated to reduce the thermal load on the detector which is cooled with a custom thermal link. The CCD baseplate assembly is shown in Figure \ref{fig:ccd_baseplate}. The CCD is model e2V CCD44-82 and will be operated at 170 K. The design of the UV-VIS detector system was described in a previous paper\cite{cosentino_vis_2018}. The progress on the detector system is reported in an accompanying paper in these proceedings.\cite{cosentino_development_2020} The full assembly is shown in Figure \ref{fig:vacuum} and is currently under test.

\section{Mounting and alignment}\label{sec:mounting}

The general alignment and integration strategy is described in a previous paper.\cite{biondi_assembly_2018} The small optics were placed on dedicated jigs which were adjusted using a CMM. Shims were used to center the elements and create gaps of specified width for the RTV adhesion pads. Examples of elements in their jigs are shown in Figures \ref{fig:grating_pics} and \ref{fig:oap_pics}.

The elements which must be aligned to good accuracy are the OAP and the camera elements. The gratings, filters, and mirrors are not sensitive to in-plane displacement, and can be positioned to better than $\pm0.1$ degrees with the CMM alone.

The OAP was placed accurately with a CMM and was considered a datum for the remainder of the feed alignment. The main alignment task was to accurately determine the focal plane of the OAP. The determination of the focal plane followed the following procedure:

\begin{enumerate}
    \item Place flat mirror, in place of OAP, normal to the design chief ray. Ensure accurate positioning with a CMM and shim accordingly.
    \item Align the autocollimator against the flat mirror to sub-arcsecond accuracy.
    \item Replace flat mirror with OAP.
    \item Place flat mirror at design focal point oriented normal to design chief ray. Due to the small size of the mirror the angular accuracy is low, but this induces only vignetting and is irrelevant for this procedure.
    \item Observe reflected cross in the autocollimator and shim the retroreflection mirror until the center of the cross is in focus.
\end{enumerate}

\begin{figure}
    \centering
    \includegraphics[width=0.45\textwidth]{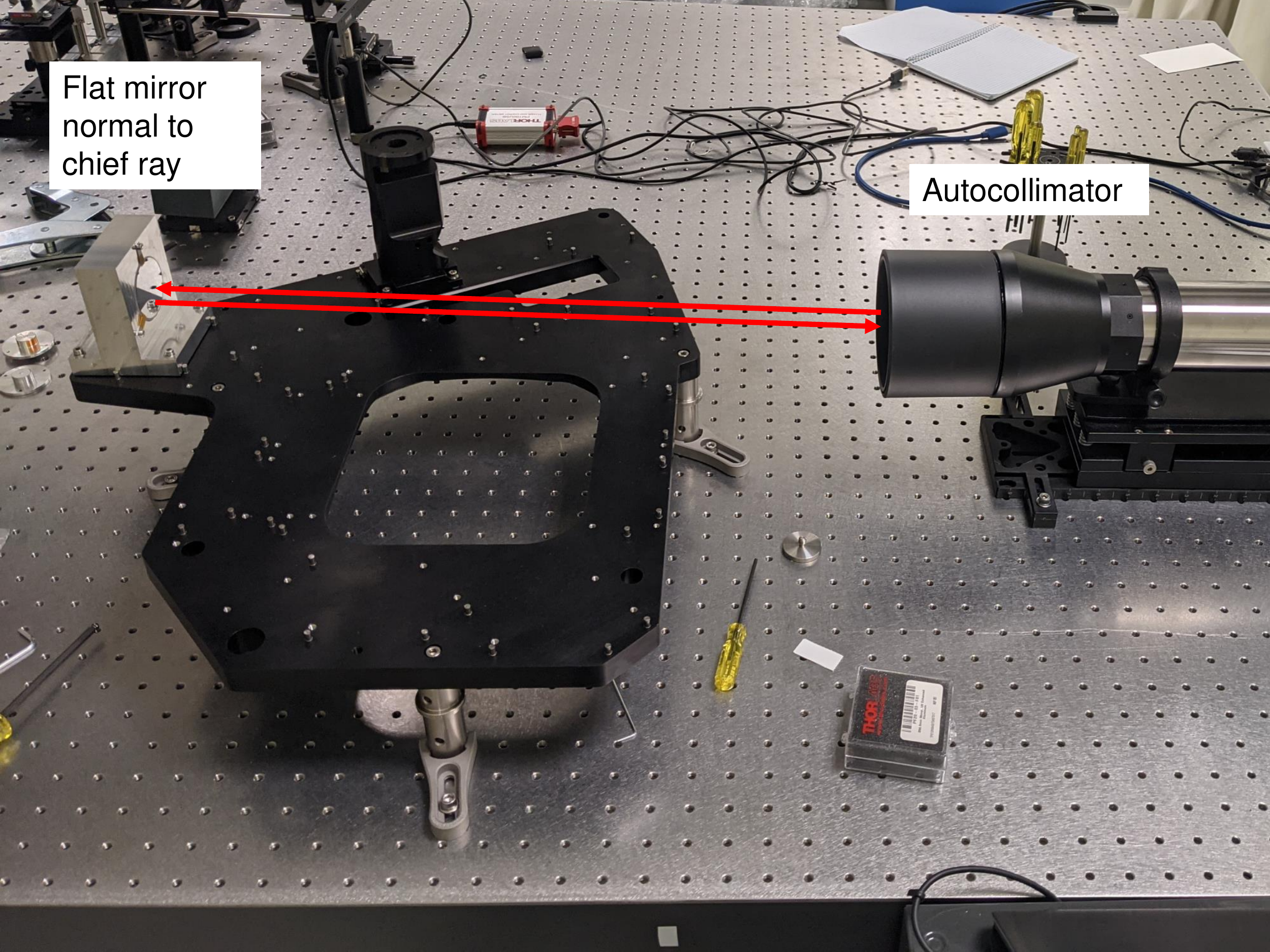}
    \includegraphics[width=0.45\textwidth]{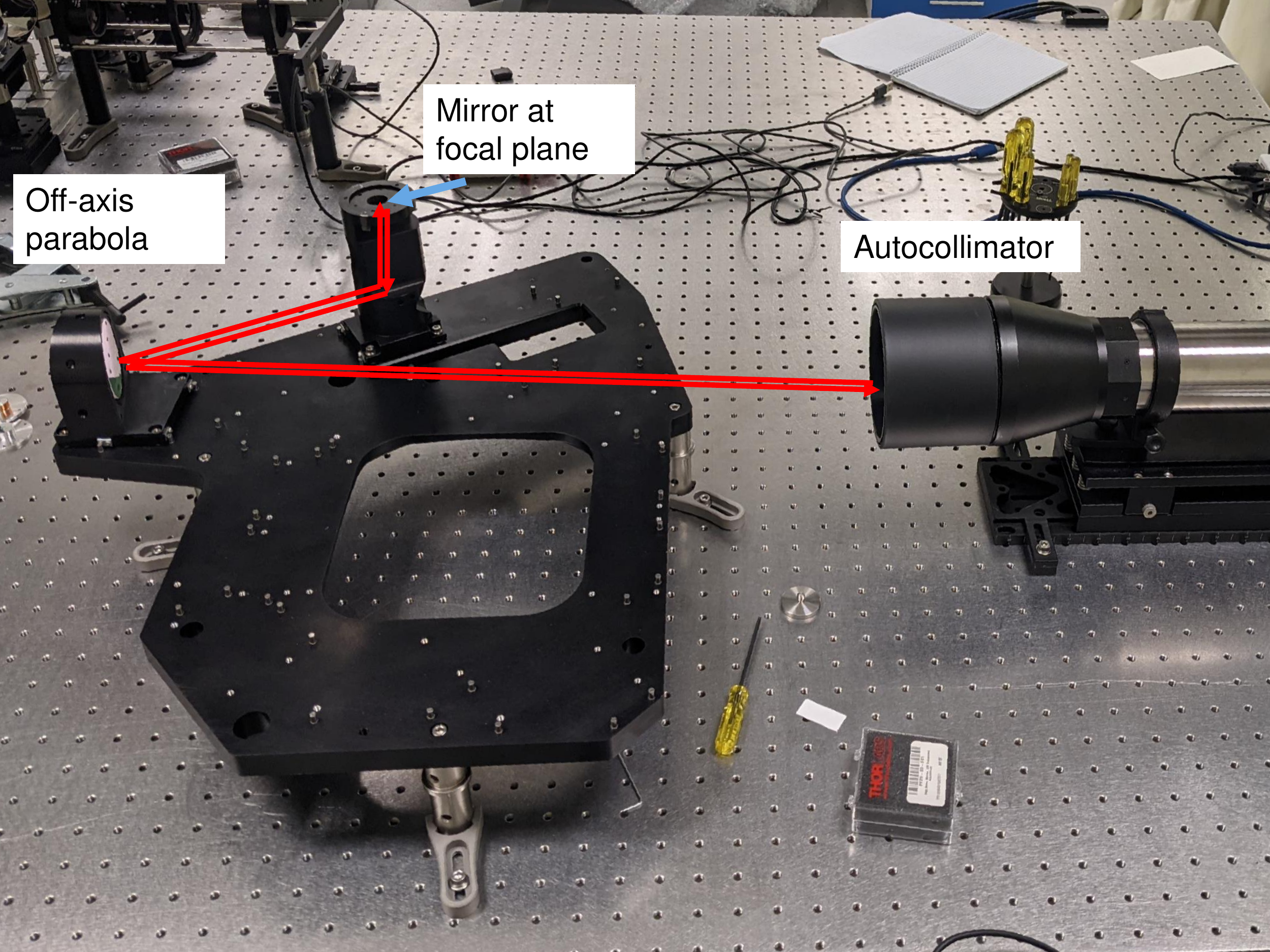}
    \caption{Alignment setup for slit position. Left, stage $1-2$: alignment of the autocollimator to the direction of the design chief ray. Right, stages $3-5$: adjustment of the focal plane mirror height to achieve a return cross that is in focus.}
    \label{fig:oap_alignment_setup}
\end{figure}

\section{Expected system performance}
\begin{figure}
    \centering
    \includegraphics[width=0.75\textwidth]{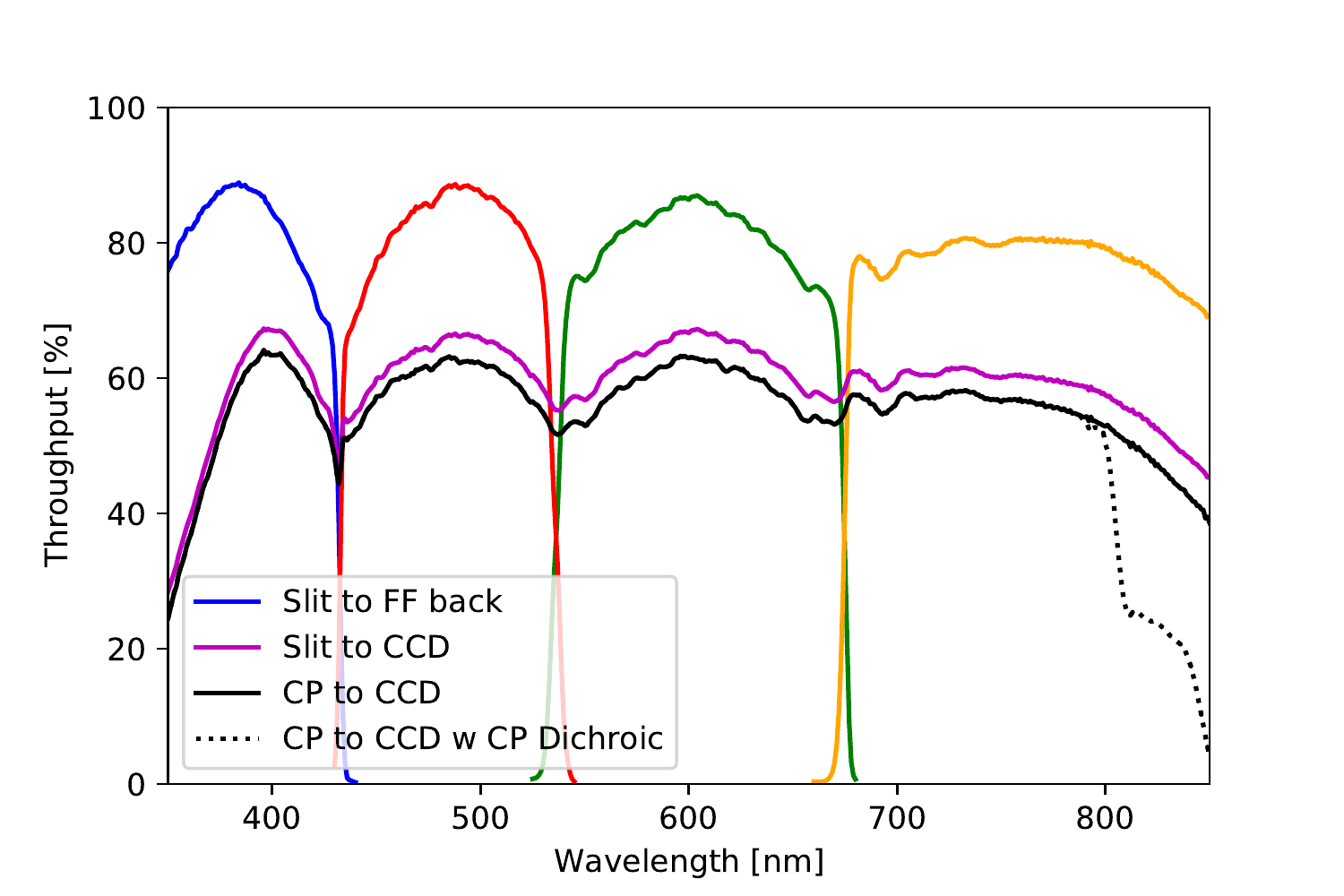}
    \caption{Total estimated efficiency, based on element as-built measurements. The average efficiency of the UV-VIS arm is 79.7\%, 59.4\%, and 55.7\% for the spectrograph optics, optics $+$ detector, and spectrograph $+$ the common path optics respectively.} 
    \label{fig:tot_eff}
\end{figure}

\begin{figure}
    \centering
    \includegraphics[width=0.75\textwidth]{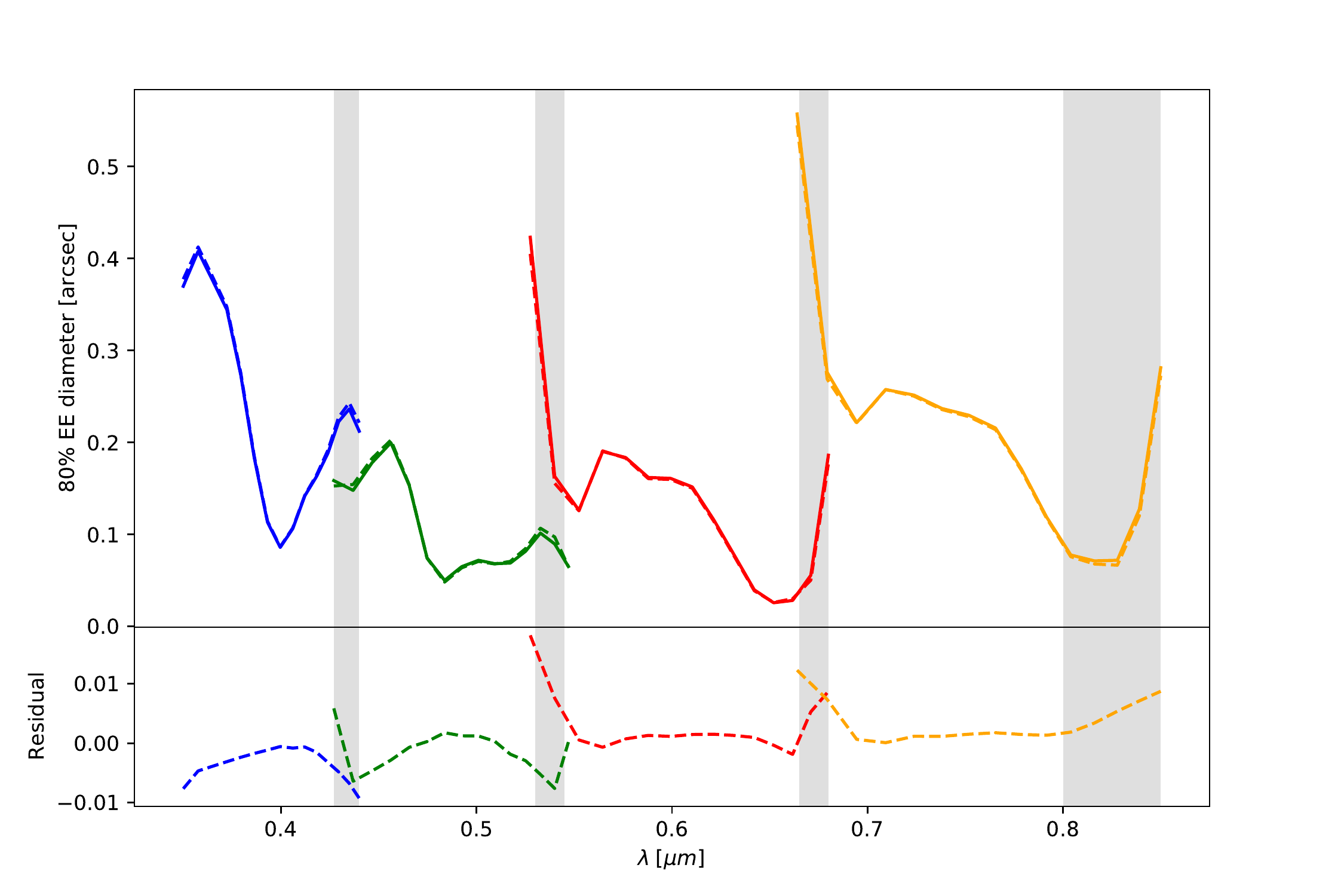}
    \caption{Estimated as-built image quality compared to design. The lower panel shows the difference between the model with the as-built element figures and the nominal design.} 
    \label{fig:asb_image_quality}
\end{figure}

The throughput of the UV-VIS arm is extremely high. Taking into account the as-built transmission and reflection measurements, the throughput is 79.7\%, 59.4\%, and 55.7\% for the spectrograph optics, optics $+$ detector, and spectrograph $+$ the common path optics respectively. Figure \ref{fig:tot_eff} shows the spectral transmission curves as estimated from the single element throughput measurements. The camera and collimator were manufactured well within tolerance, and the enslitted energy diameter is essentially nominal when taking into account the as-built prescription. The enslitted energy with the as-built prescription is shown in Figure \ref{fig:asb_image_quality}.

\section{Future activities}

The camera optics will be shipped by the end of 2020. The UV-VIS spectrograph will then be assembled and tested with an engineering detector at the Weizmann Institute in Q1 of 2021. Following successful operation at the Weizmann Institute the system will be shipped to Padua where it will be integrated with the scientific CCD and the rest of the spectrograph, with preliminary acceptance Europe currently scheduled for end of 2021.

%
%

\bibliography{report} 
\bibliographystyle{spiebib3} 

\end{document}